# Fine Structure in Magnetization of Individual Fluxoid States

A.K. Geim[†], S.V. Dubonos[*] and J.J. Palacios[o]

[†]University of Nijmegen, Toernooiveld 1, 6525 ED Nijmegen, The Netherlands

[*] Institute for Microelectronics Technology, 142432 Chernogolovka, Russia

[o]Dept. de Física Teórica de la Materia Condensada, Universidad Autónoma de Madrid, Cantoblanco, Madrid 28049, Spain

Each time a vortex enters or exits a small superconductor, a different fluxoid state develops. We have observed splitting and sharp kinks on magnetization curves of such individual states. The features are the manifestation of first and second order transitions, respectively, and reveal the existence of distinct vortex phases within a superconducting state with a fixed number of fluxoids. We show that the kinks indicate the merger of individual vortices into a single giant vortex while the splitting is attributed to transitions between different arrays of the same number of vortices.

Mesoscopic superconductors of the size comparable to the superconducting coherence length are able to accommodate only a small number of vortices before their superconductivity is destroyed and, accordingly, they have been referred to as few-fluxoid superconductors (FFS) or boxes for vortices [1-4]. How are the properties of such superconductors altered with respect to their macroscopic counterparts? Despite the obvious appeal of this simple question and all the progress made on other mesoscopic systems of similar (submicron) sizes, surprisingly little remains known about few-fluxoid superconductors. Until recently, experimental studies on this mesoscopic system were essentially limited to its behavior at the normal-superconducting phase boundary, which becomes strongly modified by the confinement. Good understanding has been achieved in this regime [5-8].

The advent of ballistic Hall micromagnetometry [9] has made it possible to investigate individual FFS *inside* the superconducting state, away from the phase boundary and the transition temperature $T_C$. Using this technique, it has been shown that a mesoscopic superconductor in a magnetic field does not exhibit the standard magnetization dependences but follows a rather dissimilar set of curves corresponding to different numbers of fluxoids (or vortices) inside. Every such vortex state exhibits its own magnetization dependence [1-4]. Furthermore, the magnetization response of FFS does not vanish at the second critical field, $H_{c2}$, but survives all the way to $H_{c3}$, emphasizing the importance of the surface superconductivity in small systems [1-10]. An essential feature of FFS is their intrinsic metastability: thermodynamically unfavorable states are very robust and can persist for indefinitely long time in a vast range of temperatures. This leads to such non-trivial consequences as, for example, a paramagnetic Meissner effect [10-12]. In more general terms, another major attraction of few-fluxoid superconductors is worth mentioning: FFS often show negligibly small pinning, presumably due to their small size [12,13]. The latter circumstance combined with the possibility to study magnetization with a resolution much better than one flux quantum, makes individual mesoscopic superconductors a valuable system for addressing unresolved problems in the physics of superconductivity.

Recent theories and numerical simulations [2-4,14-16] have reached good agreement with the previous experiments on FFS but theorists have since been raising an increasing number of new questions, challenging experimentalists. Among the most intriguing questions, in our opinion, is: how does the superconducting phase diagram of FFS differ from the macroscopic case [2,8,14-16]? In particular, the theory has predicted the possibility of second-order phase transitions *within* fluxoid states, when several Abrikosov vortices coalesce into a single giant vortex. The transition represents a mesoscopic analogue of the phase transition from bulk to surface superconductivity at $H_{c2}$ and might occur in both equilibrium and metastable situations. In this contribution, we report experimentally the existence of multiple phases within individual fluxoid states. The corresponding phase transitions can be *both* first and second order as indicated by clear jumps in magnetization and its derivative. We present an example of the superconducting phase diagram for an individual fluxoid state. The observed second-order transitions are in agreement with the behavior predicted for giant vortices [2,14,15]. The first-order transitions occur in the multi-vortex regime and are attributed to switching between superconducting states with different configurations of the same number of vortices [15,16].

We have been studying superconducting disks made of Al and Nb with diameters $d$ from 0.1 to 20 µm and thicknesses from 0.02 to 0.2 µm. The samples' magnetization was measured by Hall probes made from a high-mobility 2D electron gas and having conduction widths, $w$, between 0.5 and 20 µm. The measurement technique is described in detail in ref. [10,17]. For brevity, one can simply consider the ballistic Hall magnetometer as a fluxmeter with a square detection loop of size $w$ ($\geq d$), in the center of which a superconducting disk is placed [10]. We present our measurements in terms of the (area) magnetization $4\pi M = <B> - H$, where $<B>$ is the average magnetic field within the central area ($w \times w$), which is directly measured in the experiment [17], and $H$ is the applied field. The absolute value of $M$ depends on the filling factor $d/w$ (ref. [10,17]). In comparison to the earlier work [1,12], the sensitivity of our measurements is significantly improved (by nearly an order of magnitude; cf. Fig.1) by illuminating the Hall probes from the back through the GaAs substrate. A few seconds of infrared illumination permanently increases mobility and concentration of 2D electrons, suppressing electrical noise and allowing higher driving currents.

The studied superconducting disks are defined by electron-beam lithography and deposited by thermal evaporation. Al samples have a magnetic penetration length, $\lambda$, of the order of 100 nm at low temperatures. This value is considerably shorter than the one we could achieve in the case of Nb films and, consequently, the magnetization response for Al disks is significantly stronger, making their measurements much more accurate. In this study, we concentrate on Al. Furthermore, if the diameter of our Al disks is smaller than 0.5 µm, only the lowest fluxoid state is observed (no vortex is allowed inside such a small disk), and the situation is simple and well understood [1-3]. On the other hand, if $d$ exceeds several µm so that a disk can accommodate many dozens of vortices, we begin to observe effects of bulk pinning and, in addition, it becomes practically impossible to record magnetization curves for all fluxoid states. Therefore, Al disks with $d$ between 1 and 4 µm present an optimum system for studies of FFS.

Two examples of magnetization response for such disks are shown in Figure 1. These particular disks have $T_C \approx 1.2$ K and the critical field extrapolated to zero temperature, $H_{c3}(0)$, of about 180 Gauss, as found in magnetization measurements. Alternatively, measuring the resistivity of a macroscopic Al film evaporated simultaneously with the disks, we find the bulk critical field $H_{c2}(0) \approx 105$ G. The superconducting coherence length $\xi(0)$ is estimated to be $\approx 0.25$ µm and $\lambda(0) \approx 70$ nm [1,12], i.e. the material is a type-I superconductor ($\kappa = \lambda/\xi$

≈0.3). One has to bear in mind, however, that because of demagnetization effects, thin films in a perpendicular magnetic field behave more like type-II superconductors and are expected to exhibit vortex structures [18,2]. Note that we intentionally work with films having κ as low as possible. In fact, we can move into the true type-II regime by using less pure Al, but our tentative measurements showed no unexpected changes in overall behavior, at least, up to κ ≈2. At the same time, the reduced screening due to unavoidably larger λ for large κ, led to rapid deterioration of the experimental resolution.

Figure 1 show series of well resolved, approximately parallel magnetization curves. Each curve characterizes a different fluxoid state. These states can be described by a fluxoid (or winding) integer number $L$ that determines how many fluxoids (or vortices) are inside the disk or, alternatively, how many times the phase changes by $2\pi$ along sample's circumference [1-4]. When the magnetic field is swept continuously, the magnetization evolves along one of the solid curves in Fig. 1 until it reaches the end of this curve and jumps to the next one, belonging to another fluxoid state. Then, the process repeats itself all over again. The inset in Fig. 1a illustrates such behavior for a continuous sweep up and down. One of the important features in Fig. 1 is the existence of several fluxoid states for the same value of $H$. Apparently, only one of such states is thermodynamically stable. The ground state comprises nearly the whole low-field curve ($L =0$; the Meissner state) and segments of the other curves close to their upper (diamagnetic) ends [2-4,12,14-16]. Other states in the multiple-choice situations are metastable but can persist for many hours. This metastability is due to the inherent presence of the surface barrier [1,14,15] and leads to the hysteresis for continuous field sweeps. On the other hand, each of the fluxoid curves is completely reproducible (no hysteresis), indicating the virtual absence of bulk pinning.

The majority of fluxoid curves in Fig. 1 are bent at the diamagnetic end, which is most clearly seen for the Meissner state ($L = 0$). To our knowledge, this type of a non-linear Meissner effect has never been observed in macroscopic superconductors. In addition to the smooth curvature, there are a number of fine details that we were able to distinguish due to the high experimental resolution. A first feature to notice is additional curves in Fig. 1b that look like splitting for $L = 2, 5, 7, 8$ and $10$. As only integer fluxoid numbers are allowed, we refer to the additional fragments as substates of the corresponding fluxoid states. No such splitting has been observed for small disks with the maximum allowed $L$ less than 10 (cf. Fig. 1a). With increasing temperature, the extra segments rapidly become shorter and completely disappear above 0.8 K.

Another fine feature of the fluxoid magnetization is sharp kinks which can already be discerned in Fig. 1a for $L = 4$ and $5$ (vertical arrows). For a better view, Figs. 1b (inset) and 2a magnify two of such kinks for the larger disk. Kinks mark a rapid change in the magnetization slope and are most pronounced for $L =7$ where the slope changes by ≈40%. Kinks were observed for $L =2, 4, 5, 8, 9$ and $10$ and, in addition, we cannot exclude the presence of weak, smeared kinks for $L =3$ and 6. For another large disk studied in detail, we observed splitting and kinks also at $L = 11$, 13 and 17. Neither of our samples showed evidence of any extra features at $L = 0$ and 1 or for $L$ above $H_{c2}$.

The fluxoid substates at $L = 5, 7, 8$ and 10 in Fig. 1 persist for many hours but could only be reached on rare occasions by jumping from the corresponding ($L+1$)-states during down-sweeps. If the magnetic field is swept beyond the stability range for the substates, they usually switch to the main state that they belong to. This always happened with a jump in magnetization (first order transition). Accordingly, the substate curves in Fig. 1b do *not* touch the main fluxoid curves (see the inset and Fig. 2a). Furthermore, even if this gap is ignored, the substate curves are not a simple continuation of the main curves because they have somewhat shallower slopes. The dashed lines in the inset of Fig. 1b illustrate this.

The substate for $L =2$ turns out to be a special case. It is found at the low-field side of the main curve, in obvious contrast to the location of the other substates. The $L =2$ substate is rather unstable and falls on the main curve within several minutes at 0.3 K. This substate is, however, reached every time when sweeping the field down either from the $L =3$ state or along the main fluxoid curve. The absence of any noticeable anomaly at the transition between the solid to dotted curves in Fig. 2b suggests that this substate is the continuation of the fluxoid configuration at the diamagnetic end. For completeness, Fig. 2b also presents the superconducting phase diagram for the $L =2$ state. Note that this is the first phase diagram measured for an individual fluxoid state.

The extra features on the magnetization curves give an unambiguous indication of structural transitions in the flux distribution within fluxoid states. These transitions occur in two ways: either smoothly - in the form of a second order transition indicated by a kink (discontinuity in the magnetization derivative) - or abruptly, in the form of a first order transition (switching between the split curves). The only structural transition that is known for macroscopic superconductors and, therefore, can also be expected in mesoscopic samples, is the transition from the bulk to surface superconductivity. In bulk, the flux enclosed inside a continuous surface superconducting sheath should split at $H_{c2}$ into a number of individual vortices in a second order transition (to the best of our knowledge, this has never been observed experimentally). A mesoscopic analogue of the $H_{c2}$ transition would be the partition of a giant vortex with orbital momentum $L$ into $L$ Abrikosov vortices [8,14-16]. We attribute the observed kinks to such giant-multivortex transitions (GMT). This interpretation is directly supported by solving the full set of Ginzburg-Landau equations for our particular situation (for details, see refs. [14-16]). An example of the calculated distribution of the order parameter $\psi$ on the two sides of GMT for $L =7$ is shown in Fig. 3.

When, during a structural transition, a magnetization curve becomes steeper, it means that more flux ($<B> = H + 4\pi M$) is enclosed inside a superconductor, compared to the situation in which the curve would continue straight (see, e.g., the inset in Fig. 1b). "More flux" means that individual fluxoids inside the disk are located further away from the edge because, in such a case, a smaller amount of flux leeks outside the disk. Therefore, the direction of all the kinks (except for $L =2$) shows that, with increasing $H$, individual vortices merge into a giant vortex and not vise versa, in agreement with theory. This is also what one can generally expect because an increasing field increasingly pushes vortices away from the edge. Note that we observe no GMT either for $L = 0$ and 1 (no vortices to merge) or for large $L$ (inside the surface-superconductivity regime).

In Fig. 1, only the very last GMTs (for $L =5$ and 10, respectively) occur in the thermodynamically stable situation. Therefore, these transitions represent the equilibrium (true) $H_{c2}$ transition. Both kinks occur close to the value of $H_{c2}(0.5K) \cong 60$ G for the disks' material. All other GMTs occur in the metastable regime and represent the onset of a multivortex phase for a non-equilibrium part of the phase diagram. Note that, due to small κ, the equilibrium state of our disks is expected to be a giant vortex, even below $H_{c2}$ (ref. [14]). The observed behavior is in agreement with the theory, although the latter predicts smaller changes in the magnetization slopes and a smaller spread in the position of GMTs versus field [14-16].

Concerning the observed splitting of fluxoid curves, note first that the substates always occur in the multivortex regime where one might expect multiple possible configurations of the same number of vortices. As already mentioned, the observed substates have shallower slopes than the giant-vortex curves (Fig. 1b), which implies that individual fluxoids are more loosely distributed inside the disk compared to the case of their merger into a giant vortex. This proves that the substates are also multivortex arrays, similar to those revealed by the kinks but of different configurations. A detailed theory [15,16] confirms the possibility of such arrangements in the metastable regime and Fig. 2a plots an example of such split magnetization curves found theoretically. The other split curves in Fig. 1 can also be explained by vortices arranged in single and double rings [15]. Above $H_{c2}$, the disks are no longer in the multivortex regime, and no splitting can be expected, nor is it observed in the experiment.

The kink and splitting for $L =2$ can also be attributed to the merger of two vortices into a double-flux vortex, similar to the behavior discussed above. However, the opposite direction of this kink implies that the thermodynamically stable configuration at the diamagnetic end consists of two individual vortices, while the robust metastable configuration at low magnetizations is a giant vortex, as pictured in Fig. 2b. A transition from a giant vortex to two vortices with increasing rather than decreasing magnetic field is rather unexpected. However, there is little room for an alternative explanation, as other fluxoid configurations are not possible in the case of $L =2$. The discussed behavior is observed in three samples, which makes the feature unlikely to be defect-related. Moreover, neither an off-center pinning nor a deviation from the circular shape in our disks could explain the opposite direction of the kink.

Finally, we want to point out a rather counterintuitive (and educative) feature of the observed fluxoid transitions. Note that, for example, the transition between the two giant vortex states in Fig. 2a ($L = 6$ and $L = 7$) can occur only with a jump in magnetization (first order transition) while the splitting of a giant vortex ($L = 7$) into 7 Abrikosov vortices occurs as a second-order transition. The corresponding distributions of the density of Cooper pairs (Fig. 3) show the obvious change in symmetry for the second-order transition while changes are hardly noticeable for the transition between the two giant-vortex states. This seemingly contradicts the theory of phase transitions, which expects first order transitions to exhibit more pronounced changes in symmetry than second order transitions. This puzzle is resolved in Fig. 3 by plotting the imaginary part of $\psi$ (it can equally be its real part or the phase). It is clearly seen now that in the first-order transition there is a change from six- to seven- fold symmetry for the complex order parameter as a whole. On the other hand, the "complex" symmetry remains essentially unchanged (seven-fold) along the whole fluxoid curve $L = 7$, despite the giant-multivortex transition. This illustrates that not only the "real" distribution of Cooper pairs but also the complex (or quantum) symmetry are important in superconducting phase transitions.

In conclusion, we have observed multiple phase transitions within individual vortex states as manifested by jumps in magnetization and its derivative. We associate these transitions with the merger of vortices into a single giant vortex and switching between different (metastable) arrays of the same number of vortices. In a number of details, our interpretation has to rely on the currently developing theory and, in some cases (especially, for $L = 2$), the flux distribution remains speculative. We hope that further theoretical progress will clarify the observations and, eventually, visualization techniques will be employed to study fluxoid structures.

We gratefully acknowledge discussions with F. Peeters, V. Schweigert and V. Moshchalkov.

Figure Captions

Figure 1. Magnetizations of Al disks of diameters 1.5 (a) and 2.4 μm (b) in magnetic field along the disk axis at $T \approx 0.5$ K. The magnetometer width is $\approx 2.5$ μm. Inset in Fig. 1a shows magnetization response when the field is swept up continuously from zero to high fields and then back to zero, as the arrows indicate. Fig. 1a also allows comparison between the resolution previously achieved (inset) and the present one (figure itself). Inset in Fig. 1b magnifies one of the curves exhibiting a second order transition which corresponds to the merger of 10 vortices into a giant one. The dashed curves in the inset indicate the slope change during the transition.

Figure 2. Structural transitions for seven (a) and two (b) fluxoids of Figure 1b. The drawings schematically show the expected distribution of magnetic flux for different branches of the fluxoid curves. Inset in Fig. 2a plots a similar transition from a giant vortex to two different multivortex configurations as found by minimizing the Ginzburg-Landau functional (theoretical curves are more strongly bent at the ends as the theory allows a larger range for the fluxoid stability [15]). Inset in Fig. 2b shows the superconducting phase diagram found for the $L = 2$ state. Solid circles define the range of its existence. Open circles approximately mark the border between thermodynamically stable and metastable situations (this border is defined according to ref. 2 as the disappearance of the previous ($L = 1$) state). The dual metastable state (where the split curves are found) lies below the dashed line. All the inset's curves are guides to the eye.

Figure 3. Calculated distribution of the density of Cooper pairs (top row) and the imaginary part of the order parameter (bottom row) for six (first column) and seven fluxoids. The second and third columns correspond to giant and multivortex states for $L = 7$, respectively. White color represents minima of the functions.

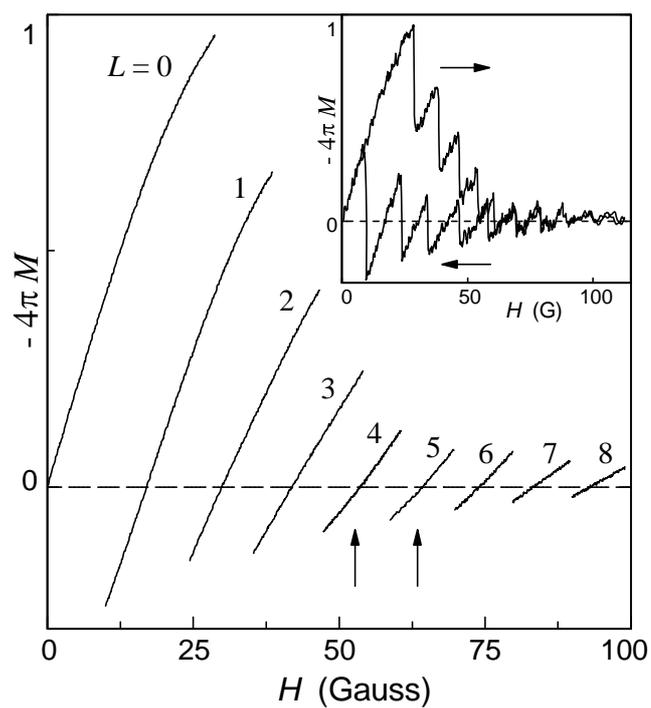

Figure 1a.

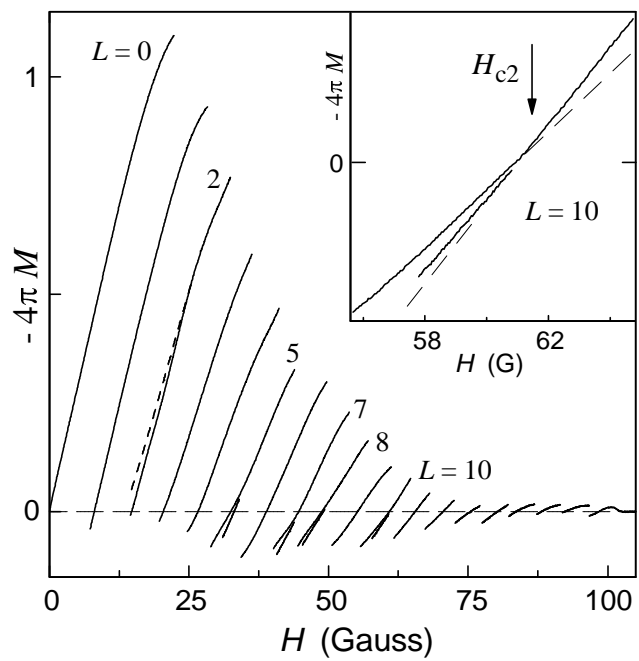

Figure 1b.

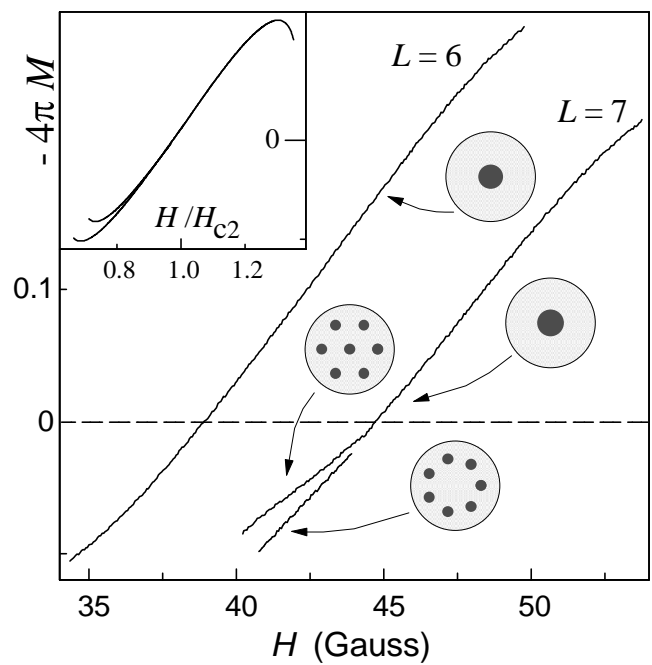

Figure 2a.

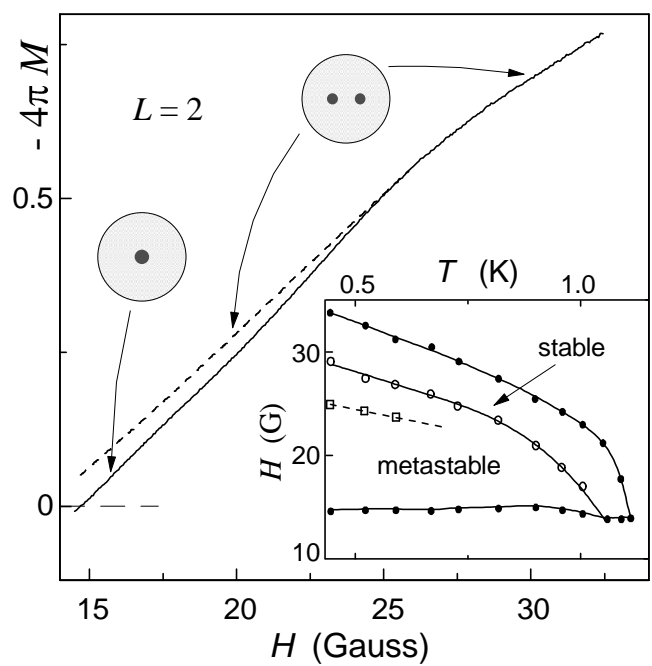

Figure 2b.

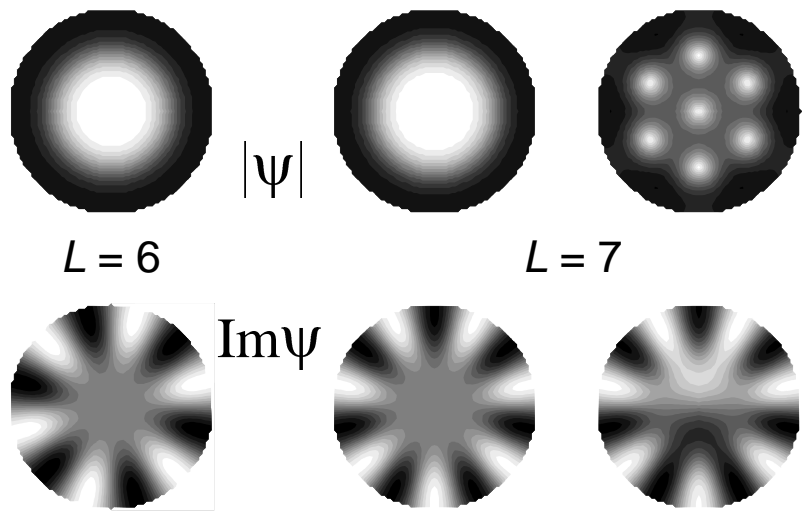

Figure 3.